\documentclass[pdflatex]{article}
 \usepackage{graphicx}
\usepackage{amsmath}
\usepackage{amssymb}
\usepackage{theorem}
\usepackage{color}

\newcommand{\A}{{\cal A}}

\newcommand{\Nb}{{\mathbb{N}}}

\newcommand{\qed}{\hfill\hbox{\rule{6pt}{6pt}}}
\newtheorem{theorem}{Theorem}

\newtheorem{col}{Corollary}

{\theorembodyfont{\rmfamily}
{\theorembodyfont{\rmfamily}

\addtolength{\topmargin}{9mm}

\interdisplaylinepenalty=2500 

\hyphenation{op-tical net-works semi-conduc-tor}

\begin{document}

\begin{center}
{\Large
The distributions of the sliding block patterns in
finite samples and the inclusion-exclusion
principles for partially ordered sets
}\bigskip

{\large
 
 Hayato Takahashi\\
  Random Data Lab.\\
                     121-0062 Tokyo Japan.\\
                        Email: hayato.takahashi@ieee.org}
                        
\end{center}

\begin{abstract}
 The sliding block patterns are the random variables that count the number of the appearance of words in finite samples.
In this paper we show a new formula of  the distributions of sliding block patterns for Bernoulli processes with finite alphabet.
In particular we show a new inclusion-exclusion principle on partially ordered sets with multivariate generating function,   and 
give  a simple formula of the distribution of the sliding block patterns with generating functions. 
We also show the formula of higher moments of the sliding block patterns.
By comparing the powers of tests,  we show the significant performance of the sliding block patterns tests.
We  show that the sliding block patterns tests  reject the BSD Library RNG with p-value almost zero.
 \end{abstract}

\section{Introduction}
We study the word counting problem, i.e., the number of appearance of words in finite samples.
For example let us consider the word \(01\) and the finite sample \(10101101\).
Then the word \(01\) appears three times in the sample and there are seven runs in  \(10101101\).
Note that the number of appearance of \(01\) is almost same to  twice the number of runs.
Statistical tests based on the number of appearance of the words are considered to be a generalization of  the run tests.

Let \(X\in A^n\) with finite alphabet \(A\) and \(w\in  A^*\). Let \(|w|\) be the length of the word \(w\).
We consider the following random variable,
\begin{equation}\label{sliding-block-A}
N_w:= \sum_{i=1}^n I_{X_i^{i+|w|-1}=w}\text{ if }X_i^{i+|w|-1}=w\text{ else }0,
\end{equation}
where \(I_{X_i^{i+|w|-1}=w}=1\).
We also consider the vector of random variables \((N_{w_1}, \ldots, N_{w_l})\) and call them {\it sliding block patterns} or {\it suffix tree}.
We call statistical tests based on  sliding block patterns {\it sliding block patterns tests}.
In this paper we study  sliding block patterns tests with non-overlapping increasing multiple words (Theorem~\ref{th-main}).

The statistics of the sliding block patterns plays important role in information theory, ergodic theory, and DNA analysis, see \cite{JaquetandSzpankowski}.
Ergodic theory shows the existence empirical distributions of the sliding block patterns in the limit with probability one.
Data compression scheme LZ~77 is based on the suffix tree. LZ~78 scheme is based on return time but it is closely related to suffix tree \cite{shields96}.
These data compression schemes are also applied to nonparametric statistics \cite{LV2008}.

In order to carry out statistical tests with sliding block patterns, we need to derive the distributions of the sliding block patterns with respect to null hypotheses. 

It is well known that Monte Carlo simulation may generate a false distributions.
In Section~\ref{sec-pseudo}  we show that  Monte Carlo simulation with the BSD Library pseudo random number generator {\bf random} (BSD RNG) and 
Mersenne twister pseudo random number generator (MT RNG).  We show that BSD RNG
computes  a false distribution. 
The Monte Carlo simulation is itself considered to be a statistical test for pseudo random numbers. In order to avoid a circular argument, we need to derive
the distribution with mathematically assured manner.

The distributions of sliding block patterns have been shown via generating functions, see \cite{{Guibas81},{RegnierSpankowski98},{JaquetandSzpankowski},{Goulden83},{Bassinoetal2011},{Flajolet2009}}.
R\'{e}gnier and Szpankowski \cite{RegnierSpankowski98} derived  generating functions of the sliding block patterns 
in a finite sample.
In  Goulden and Jackson \cite{Goulden83} and Bassino et al \cite{Bassinoetal2011}, they obtained the distribution of  the sliding block patterns  with 
generating functions  and inclusion-exclusion method.
The advantage of the method of Bassino  \cite{Bassinoetal2011} is that the  formula of generating functions 
are significantly  simplified in combination with  inclusion-exclusion method. 
The formula of the generating function of the distribution of the sliding block patterns  in \cite{{Guibas81},{RegnierSpankowski98},{JaquetandSzpankowski},{Goulden83},{Bassinoetal2011},{Flajolet2009}} are based on 
the induction of sample size.

In this paper we show the distributions of sliding block patterns for Bernoulli processes with finite alphabet, which is not based on the induction on sample size.
We show a new inclusion-exclusion formula in multivariate generating function form on partially ordered sets,
and show a simpler expression of generating functions of the 
number of pattern occurrences in finite samples. 

We say that a word \(w\) is overlapping if there is a word \(x\) with \(|x|<2|w|\) and \(w\) appears in \(x\) at least 2 times, otherwise
 \(w\) is called non-overlapping.  
For example \(10\) is non-overlapping and \(11\) is overlapping, i.e., \(11\) appears 2 times in \(111\).
We write \(x\sqsubset y\) if \(x\) is a prefix of \(y\) and \(x\ne y\).
\begin{theorem}\label{th-main}
Let \(P\) be an i.i.d.~process of fixed sample size \(n\) of finite alphabet. 
Let \(s_1\sqsubset s_2\sqsubset\cdots \sqsubset s_l\) be an increasing  non--overlapping words of finite alphabet, i.e., \(s_i\) is a prefix of \(s_j\)  and  \(m_i<m_j\), where \(m_i\) is the length of \(s_i\), for all  \(i<j\).
Let \(P(s_i)\) be the probability of \(s_i\) for \(i=1,\ldots, l\).
Let 
\begin{align}
&A(k_1,\ldots,k_l)=\dbinom{n-\sum_i m_i k_i + \sum_i k_i}{k_1,\ldots, k_l}\prod_{i=1}^l P^{k_i}(s_i),\nonumber\\
&B(k_1,\ldots,k_l)=P(\sum_{i=1}^n I_{X_i^{i+m_i-1}=s_j}=k_j,\ j=1,\ldots,l),\label{eq-A}\\
&F_A(z_1,\ldots,z_l)=\sum_{k_1,\ldots,k_l}A(k_1,\ldots,k_l)z^{k_1}\cdots z^{k_l},\text{ and}\nonumber\\
&F_B(z_1,\ldots,z_l)=\sum_{k_1,\ldots,k_l}B(k_1,\ldots,k_l)z^{k_1}\cdots z^{k_l}.\nonumber
\end{align}
 Then
\[
F_A(z_1,z_2,\ldots,z_l)=F_B(z_1+1, z_1+z_2+1,\ldots, z_1+\cdots+z_l+1).
\]
Or equivalently,
\[
F_A(y_1-1,y_2-y_1,\ldots, y_l-y_{l-1})=F_B(y_1,y_2,\ldots, y_l),\]
where \(y_i=z_1+\cdots+z_i+1\) for  \(i=1,\dots,l\). 

\(B(k_1,\ldots,k_l)\) is the coefficient of the 
\(\prod_{i=1}^l y_i^{k_i}\) in \(F_A(y_1-1,y_2-y_1,\ldots, y_l-y_{l-1})\) for all \(k_1,\ldots, k_l\).
\end{theorem}
It is known that in the case of one variable  \cite{HSWilf2006} or the {\it multi-variate disjoint events} \cite{Bassinoetal2011,Goulden83,Guibas81}, 
inclusion-exclusion formula expressed via generating functions as \(F_A(z_1,\ldots, z_l)=F_B(z_1+1,\ldots,z_l+1)\), where \(F_A\) is  a generation function for non-sieved events
and \(F_B\) is a generating function for sieved events.
Theorem~\ref{th-main} shows a new inclusion-exclusion formula for partially ordered sets.

With slight modification of Theorem~1, we can compute the number of the occurrence of the overlapping increasing words.
For example, let us consider increasing self-overlapping words  11, 111, 1111 and  the number of  their occurrences.
Let  011, 0111, 01111 then these words are  increasing non-self-overlapping words.
The number of occurrences 11, 111, 1111 in sample of length \(n\) is equivalent to  the number of occurrences 011, 0111, 01111 in sample of length \(n+1\) that starts with 0.

In \cite{RegnierSpankowski98}, expectation, variance, and CLTs (central limit theorems) for the sliding block pattern are shown. We show the higher moments for non-overlapping words.
\begin{theorem}\label{th-moments}
Let \(w\) be a non-overlapping pattern.
\[\forall t\ E(N^t_w)=\sum_{s=1}^{\min \{T,t\}} A_{t,s}\dbinom{n-s|w|+s}{s}P^s(w).\]
\[A_{t,s}=\sum_r \dbinom{s}{r}r^t(-1)^{s-r},\  T=\max \{t\in\Nb\ \vert\  n-t|w|\geq 0\}.\]
\end{theorem}
In the above theorem, \(A_{t,s}\) is the number of surjective functions from \(\{1,2,\ldots,t\}\to\{1,2,\ldots,s\}\) for \(t,s\in\Nb\), see \cite{Riordan}.

The preliminary versions of the paper have been presented at \cite{{takahashi2018MSJ},{takahashi2018ergod},{takahashi2018ProbSympo},{takahashi2018SITA}}.

\section{Sparse Patterns}
In this section, we expand the notion of non-overlapping patterns to {\it sparse patterns}.
First we expand the notion of non-overlapping.  Two words  \(w_1\) and \(w_2\) are called non-overlapping if 
there is no word \(x\) such that \(|x|<|w_1|+|w_2|\) and  \(w_1\) and \(w_2\) appear in \(x\).
For example,  the words 00101 and 00111 are non-overlapping. 
A set \(S\) of words is called non-overlapping if \(w_1\) and \(w_2\) are non-overlapping for all \(w_1,w_2\in S\) including the case \(w_1=w_2\).
We introduce the symbol  \(?\) to represent arbitrary symbols. 
Let \(\A\) be the alphabet. A word consists of extended alphabet  \(\A\cup\{?\}\) is called sparse pattern.
We say \(w'\) is a realization of the sparse pattern \(w\) if \(w'\) consists of \(\A\) and coincides with \(w\) except for the symbol \(?\).
A sparse pattern is called non-overlapping if the set of the realization is non-overlapping.
For example \(001? 1\) is  a non-overlapping sparse pattern and its realizations are  00101 and 00111 with \(\A=\{0,1\}\).

We can find many non-overlapping sparse patterns.
For example, each sparse pattern \(0^{m+1}(1?^m)^n 1\) is non-overlapping for all \(n,m\).
Here \(w^m\) is the \(m\) times concatenation of the word \(w\).
For example, \(0^3(1?^2 )^21=0001?? 1?? 1\).

The probability of sparse pattern \(w\) is defined by
\[P(w)=\sum_{w'\text{: realization of }w}P(w').\]
We write \(w_1\sqsubset w_2\) for two sparse patterns if \(w_1\) is a prefix of \(w_2\) with the alphabet  \(\A\cup\{?\}\).
Theorem~\ref{th-main} holds for sparse words. 
\begin{col}
Theorem~\ref{th-main} holds for non-overlapping  increasing sparse patterns. 
\end{col}

The advantage of the sparse patterns is that  we can construct  large size sparse patterns that have large probabilities, which is useful in statistical tests of pseudo random numbers 
in a large sample size. 

\section{Experiments on power function of sliding block patterns tests}
In  \cite{RegnierSpankowski98}, it is shown that central limit theorem holds for sliding block patterns,
\[ P(\frac{N_w-E(N_w)}{\sqrt{V_w}}<x)\to \frac{1}{\sqrt{2\pi}}\int^x_{-\infty} e^{-\frac{1}{2}x^2} dx,\]
where \(w\) is  non-overlapping pattern,
\begin{align}\label{exp-variance}
&E(N_w)=(n-|w|+1)P(w) \text{ and} \nonumber\\
&V(N_w)=E(N_w)+(n-2|w|+2)(n-2|w|+1)P^2(w)-E^2(N_w).
\end{align}

Let
\[N'_w:= \sum_{i=1}^{\lfloor n/|w|\rfloor} I_{X_{i*|w|}^{(i+1)*|w|-1}=w}.\]
\(N'_w\) obeys binomial law if the process is i.i.d.
We call \(N'_w\) {\it block-wise sampling}. 

As an application of CLT approximation, we compare power functions of sliding block sampling \(N_w\) and block-wise sampling \(N'_w\).

We consider the following test for sliding block patterns:
We write \(E_\theta=E(N_w)\) and  \(V_\theta=V(N_w)\) if \(P(w)=\theta\).
Null hypothesis: \(P(w)=\theta^*\) vs alternative hypothesis \(P(w)<\theta^*\).
Reject null hypothesis if and only if \(N_w<E_{\theta^*} -5\sqrt{V_{\theta^*}}\).
The likelihood of the critical region is called power function, i.e.,  \(Pow(\theta):=P_\theta(N_w<E_{\theta^*} -5\sqrt{V_{\theta^*}})\) for \(\theta\leq \theta^*\).

We construct a test for block-wise sampling:
Null hypothesis: \(P(w)=\theta^*\) vs alternative hypothesis \(P(w)<\theta^*\).
Reject null hypothesis if and only if \(N'_w<E'_{\theta^*} -5\sqrt{V'_{\theta^*}}\), where \(E'_\theta = \lfloor n/|w|\rfloor \theta\) and \(V'_\theta = \lfloor n/|w|\rfloor \theta (1-\theta)\).

The following table shows powers of  sliding block patterns tests and block wise sampling at \(\theta=0.2, 0.18, 0.16\) under the condition that 
alphabet size is 2 (binary data), \(\theta^*=0.25, |w|=2,\text{ and } n=500\).

\begin{center}
  \begin{tabular}{@{} cccc @{}}
    \hline
\(\theta\) &0.2& 0.18& 0.16 \\ 
    \hline
Power of Sliding block &0.316007 &0.860057& 0.995681 \\ 
Power of  Block wise  & 0.000295 & 0.002939 & 0.021481\\ 
    \hline
  \end{tabular}
\end{center}

\begin{figure}[h]
\includegraphics[width=8cm]{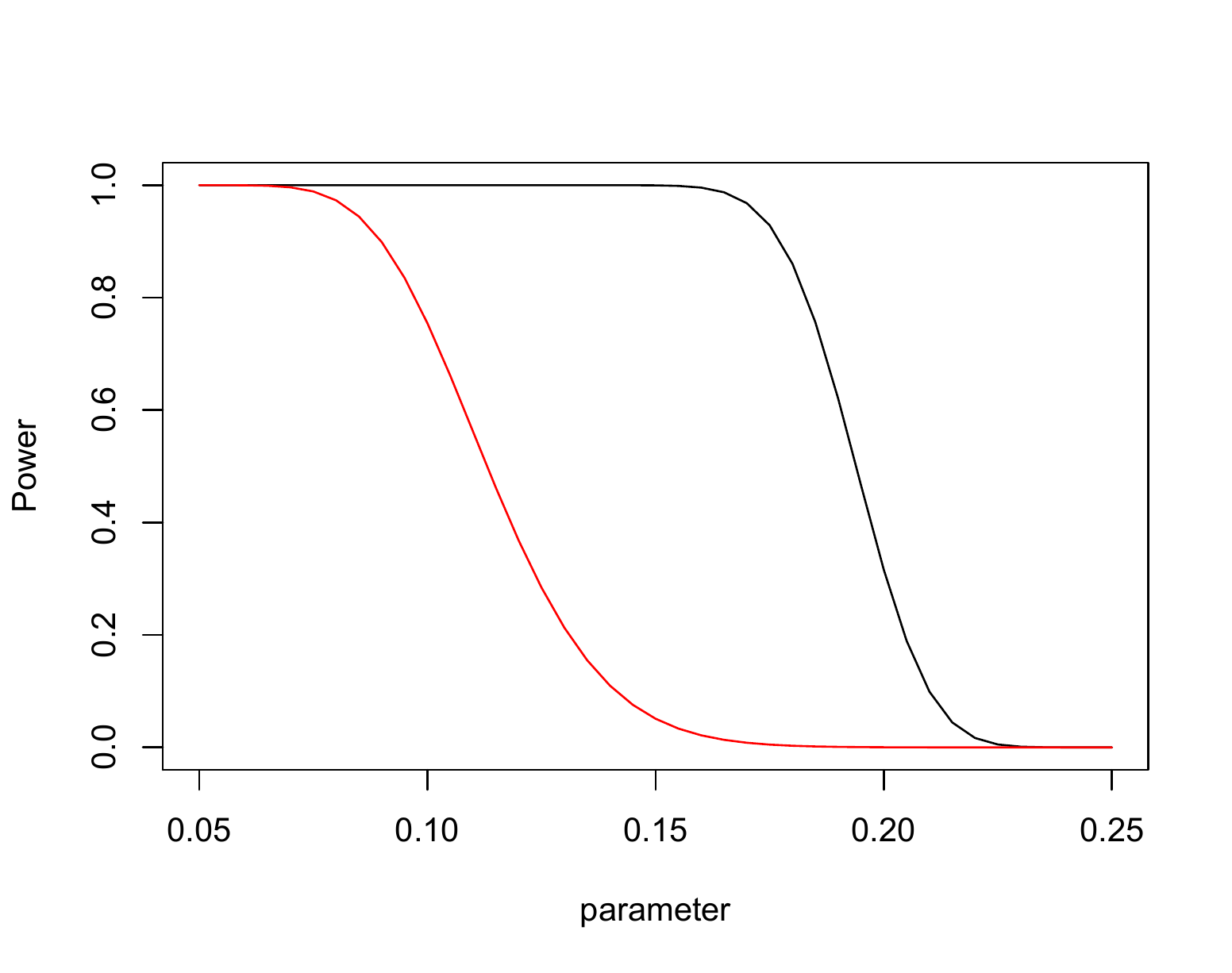}
\caption{Comparison of power functions: sliding block test (black line) vs block-wise test (red line).
\(P(w)< 0.25\) vs \(P(w)=0.25\) (null hypothesis).  \(|w|=2, n=500\).
}\label{fig-A}
\end{figure}

Figure~\ref{fig-A}  shows the graph of power functions for sliding block patterns test and block-wise sampling.

\section{Experiments on statistical tests for pseudo random numbers}\label{sec-pseudo}
In  \cite{NIST2010}, a battery of statistical tests for pseudo random number generators are proposed, and 
chi-square test is recommended to test the pseudo random numbers with
 sliding block patterns \(N_w\)   and non-overlapping word \(w\).
 Expectation and variance of \(N_w\) are given in (\ref{exp-variance}).

\begin{figure}[ht]
\includegraphics[width=6cm]{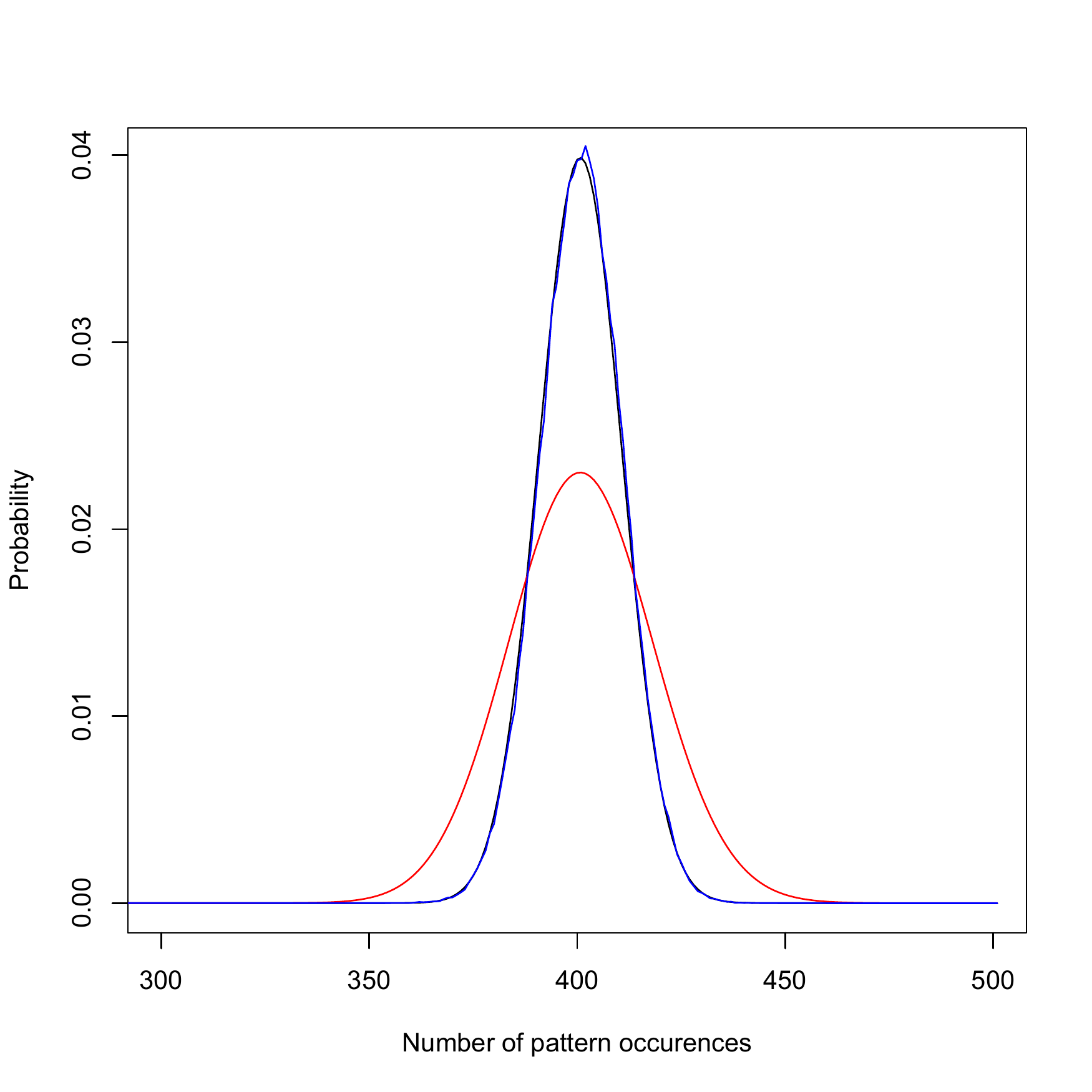}
\includegraphics[width=6cm]{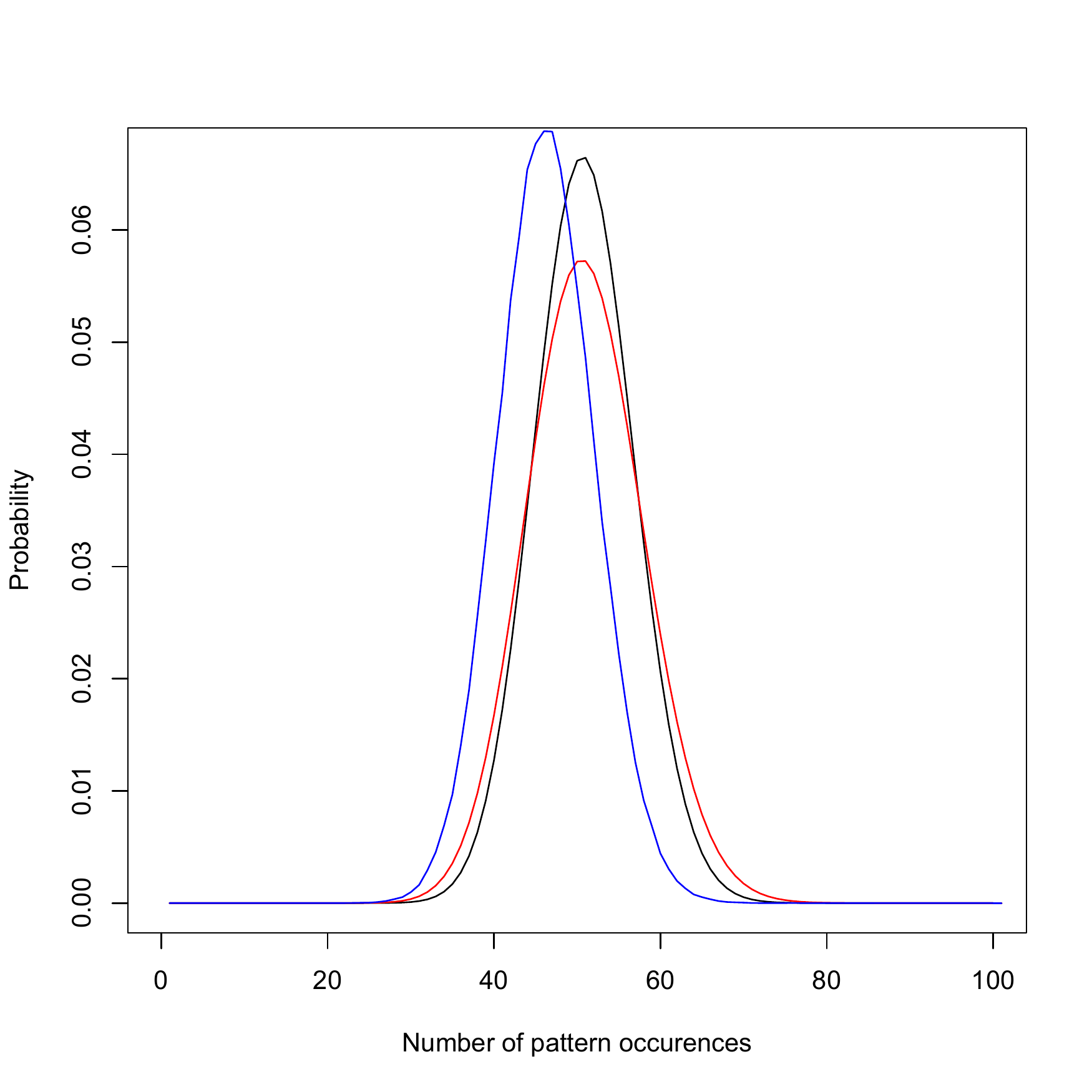}
\caption{Comparison of distributions: the left graph shows the distributions   for \(w=10\) and   \(n= 1600\) and the right graph shows the distributions for  \(w=11110\) and  \(n= 1600\) in (\ref{sliding-block-A}).
Black, red, and blue lines show true distribution, binomial, and Monte Carlo simulation, respectively.}\label{fig-dist}
\end{figure}

In this section, we  apply Theorem~\ref{th-main} and Kolmogorov Smirnov test to pseudo random number generators.
Fix sample size \(n= 1600\) in (\ref{sliding-block-A}) and null hypothesis \(P\) be fair coin flipping.
We compute the  following three distributions for \(w=10\)  and \(11110\) in Figure~\ref{fig-dist}.
\bigskip

\noindent
1. true distributions of \(N_w\),\\
2. binomial distributions\\
 \( \binom{n}{k}p^k(1-p)^{n-k},\ p=2^{-|w|},\ k=1,\ldots,n\), and\\
3. empirical distributions of \(\sum_{i=1}^n I_{X_i^{i+|w|-1}=w}\) generated by Monte Carlo method with BSD RNG {\bf random}, 
200000 times iteration of random sampling.

Due to the linearity of the expectation, the expectation of binomial distribution is \(pn\), which is almost same to the expectation of \(N_w\).
However the random variables of sliding block patterns have strong correlations even if the process is i.i.d.
For example,  if a non-overlapping pattern has occurred at some position, then the same non-overlapping pattern do not occur in the next position.

From the graphs of distributions, we see that binomial distributions has large  variance compared to the true distributions.
This is because, in the binomial model,  the correlations of patterns are not considered.   
We see that the binomial model approximations of the distributions of the sliding block patterns are  not appropriate. 

Figure~\ref{fig-dist} shows that the empirical distributions (Monte Carlo method)  is different from the true distribution.
We see that BSD RNG {\bf random} cannot simulate the sliding block patterns correctly.

From the experiment for \(w=11110\) we have obtained
\[\sup_{0\leq x<\infty}| F_t(x)-F(x)|=0.302073,\]
where  \(F_t(x)\) is the empirical cumulative distribution generated by Monte Carlo method with BSD RNG {\bf random}
with  \(t=200000\) times random sampling. 
\(F(x)\) is the cumulative distributions of  (\ref{eq-A}).
From Kolmogorov-Smirnov theorem we have the p-value
\[P(\sup_{0\leq x<\infty}| F_t(x)-F(x)|\geq 0.302513 )\approx 0,\]
where \(P\) is the fair coin flipping (null hypothesis).
The sliding block patterns tests reject BSD RNG {\bf random}.
The p-values of the Kolmogorov-Smirnov test for BSD RNG  with \(w=10,\ w=11110,\ t=200000\) and \(n=1600\) are  summarized in the following table.

\begin{center}
  \begin{tabular}{@{} cccc @{}}
    \hline
BSD RNG &w=10& w=11110 \\ 
    \hline
\(\sup_{0\leq x<\infty}| F_t(x)-F(x)|\) &0.012216 & 0.302073\\ 
p-value  & 0 & 0\\ 
    \hline
  \end{tabular}
\end{center}

The sliding block patterns tests  do not reject MT RNG \cite{MT98} under the same condition above.
The p-values of the Kolmogorov-Smirnov test for MT  RNG  with \(w=10,\ w=11110,\ t=200000\) and \(n=1600\)  are  summarized in the following table.

\begin{center}
  \begin{tabular}{@{} cccc @{}}
    \hline
MT RNG &w=10& w=11110 \\ 
    \hline
\(\sup_{0\leq x<\infty}| F_t(x)-F(x)|\) & 0.001376& 0.001409\\ 
p-value  &0.843306  & 0.822066\\ 
    \hline
  \end{tabular}
\end{center}

\section{Proofs}
Proof of Theorem~\ref{th-main})
For simplicity we give a proof for the case of two non-overlapping  words \(w_1\sqsubset w_2\) in Theorem~\ref{th-main}.
The proof of the general case is similar. 
Let \(m_1=|w_1|\) and \(m_2=|w_2|\). 
Since  \(w_1\) and \(w_2\) are non-overlapping,
the number of possible allocations such that \(k_1\) times appearance of \(w_1\) and \(k_2\) times appearance of \(w_2\)  in the string of length \(n\)  is
\[\dbinom{n-m_1k_1-m_2k_2+k_1+k_2}{k_1,k_2}.\]
This is because if we replace  each \(w_1\) and \(w_2\) with additional two extra  symbols \(\alpha, \beta\) in the string of length \(n\) then the problem  reduces to 
choosing  \(k_1\)  \(\alpha\)'s  and \(k_2\)  \(\beta\)'s among string of length  \(n-m_1k_1-m_2k_2+k_1+k_2\).
In the above equation, we do not count the appearance of \(w_1\) in \(w_2\).
Let
\[
A(k_1,k_2)=\dbinom{n-m_1k_1-m_2k_2+k_1+k_2}{k_1,k_2}P^{k_1}(w_1)P^{k_2}(w_2).
\]
\(A\) is not the probability that   \(k_1\) \(w_1\)'s  and \(k_2\) \(w_2\)'s  occurrence in  the string, since 
we allow any words in the remaining place of the string except for the appearance of  \(w_1\) and \(w_2\).
For example \(A\) may count the event that 
 \(w_1\) and \(w_2\) appear more than \(k_1\) and \(k_2\) times.
Let \(B(t_1,t_2)\) be the probability that  non-overlapping words \(w_1\) and \(w_2\) appear \(k_1\) and \(k_2\) times respectively.
We have the following identity,
\begin{align}\label{eq-proof-A}
A(k_1,k_2)=&\sum_{k_2\leq t_2, \ k_1+k_2\leq t_1+t_2}  B(t_1,t_2)\dbinom{t_2}{k_2} 
\sum_{0\leq s\leq t_2-k_2}\dbinom{t_2-k_2}{s}\dbinom{t_1}{k_1-s}.
\end{align}
Let \(F_A(z_1,z_2):=\sum_{k_1,k_2}A(k_1,k_2) z^{k_1}z^{k_2}\) and 
\(F_B(z_1,z_2):=\sum_{k_1,k_2}B(k_1,k_2) z^{k_1}z^{k_2}\) be generating functions for \(A\) and \(B\) respectively.
From (\ref{eq-proof-A}), we have
\begin{align*}
F_A(z_1,z_2)=&\sum_{k_1,k_2}z_1^{k_1}z_2^{k_2} \sum_{k_2\leq t_2, \ k_1+k_2\leq t_1+t_2} B(t_1,t_2)
\dbinom{t_2}{k_2}\sum_{0\leq s\leq t_2-k_2}\dbinom{t_2-k_2}{s}\dbinom{t_1}{k_1-s}\\
=&\sum_{t_1,t_2}B(t_1,t_2)\sum_{k_2\leq t_2}\dbinom{t_2}{k_2}z_2^{k_2}  
\sum_{0\leq s\leq t_2-k_2,\ 0\leq k_1-s\leq t_1}\dbinom{t_2-k_2}{s}\dbinom{t_1}{k_1-s}z_1^{k_1}\\
=&\sum_{t_1,t_2}B(t_1,t_2)\sum_{k_2\leq t_2} \dbinom{t_2}{k_2}z_2^{k_2} (z_1+1)^{t_1+t_2-k_2}\\
=&\sum_{t_1,t_2}B(t_1,t_2)(z_1+1)^{t_1+t_2}(\frac{z_2}{z_1+1}+1)^{t_2}\\
=&F_B(z_1+1,z_1+z_2+1).
\end{align*}
In the above  second equality, we changed the order of summation of variables.
The latter part of the theorem is obvious. \qed
\bigskip

Proof of Theorem~\ref{th-moments})
For simplicity let
\(Y_i=I_{X_i^{i+|w-1}}=w\).
Since \(w\) is non-overlapping, 
\(Y_i Y_j =Y_i\) if \(i=j\). \(Y_i Y_j =0\)  if \(\{i,i+1,\ldots,i+|w|-1\}\cap \{j,j+1,\ldots,j+|w|-1\}\ne\emptyset\). 
We say that \(\{i,i+1,\ldots,i+|w|-1\}\) is the coordinate of \(Y_i\).
We say that a subset of \(\{Y_i\}_{1\leq i\leq n-|w|+1}\) is disjoint if their coordinate are disjoint.

Let \(Y_{i,j}=Y_i\) for all \(1\leq j\leq t\).
Then  \((\sum_i Y_i)^t=\prod_{j=1}^t\sum_i Y_{i,j}=\sum_i \prod_{j=1}^t Y_{i,j}\).
Note that \(E( \prod_{j=1}^t Y_{i,j})=P^s(w)\) if and only if  there is a disjoint set  \(Y_{n(j)}, 1\leq j\leq s\) such that  \(\prod_{j=1}^t Y_{i,j}=\prod_{j=1}^s Y_{n(j)}\).

Observe that the number of possible combination of  disjoint sets of  \(Y_{n(j)}, 1\leq j\leq s\) such that  \(\prod_{j=1}^t Y_{i,j}=\prod_{j=1}^s Y_{n(j)}\) is
\(A_{t,s}\dbinom{n-s|w|+s}{s}\).  Note that there is no disjoint sets of  \(Y_{n(j)}, 1\leq j\leq s\)  if \(s>\max \{t\in\Nb\ \vert\  n-t|w|\geq 0\}\).
From  the linearity of the expectation, we have the theorem.

 \qed

\section*{Acknowledgment}

The author thanks for a helpful discussion with Prof. S.~Akiyama and Prof.~M.~Hirayama at Tsukuba University. This work was supported by the Research Institute for Mathematical Sciences, an International Joint Usage/Research Center located in Kyoto University, and Ergodic theory and related fields in Osaka University.

\end{document}